\documentclass[aps,prc,reprint,showpacs,floatfix]{revtex4-1}

\usepackage{graphicx}
\usepackage{dcolumn}
\usepackage{bm}

\begin{document}


\title{Exploring the multi-humped fission barrier of $^{238}$U via sub-barrier photofission}

\author{L. Csige$^{1,2,4}$}
\author{D.M. Filipescu$^3$}
\author{T. Glodariu$^3$}
\author{J. Guly\'as$^4$}
\author{M.M. G\"unther$^{5}$}
\author{D. Habs$^{5}$}
\author{H.J. Karwowski$^{6,7}$}
\author{A. Krasznahorkay$^4$}
\author{G.C. Rich$^{6,7}$}
\author{M. Sin$^3$}
\author{L. Stroe$^3$}
\author{O. Tesileanu$^3$}
\author{P.G. Thirolf$^{1}$}
\affiliation{
$^1$Ludwig-Maximilians-Universit\"at M\"unchen, D-85748 Garching, Germany \\
$^2$Excellence Cluster Universe, D-85748 Garching, Germany\\
$^3$H. Hulubei National Institute of Physics and Nuclear Engineering, Bucharest, Romania\\
$^4$Institute of Nuclear Research of the Hungarian Academy of Sciences (ATOMKI),
Post Office Box 51, H-4001 Debrecen, Hungary \\
$^5$Max-Planck-Institute for Quantum Optics, D-85748 Garching, Germany \\
$^6$Department of Physics and Astronomy, University of North Carolina, Chapel
Hill, NC 27599, USA \\
$^7$Triangle Universities Nuclear Laboratory (TUNL), Durham, NC 27708, USA}
\date{\today}

\begin{abstract}

The photofission cross-section of $^{238}$U was measured at sub-barrier energies
as a function of the $\gamma$-ray energy using, for the first time, a
monochromatic, high-brilliance, Compton-backscattered $\gamma$-ray beam. The
experiment was performed at the High Intensity $\gamma$-ray Source (HI$\gamma$S)
facility at beam energies between $E_{\gamma}$=4.7 MeV and 6.0 MeV 
and with $\sim3\%$ energy resolution. Indications of transmission resonances
have been observed at $\gamma$-ray beam energies of $E_{\gamma}$=5.1 MeV and 
5.6 MeV with moderate amplitudes. The triple-humped fission barrier parameters
of $^{238}$U have been determined by fitting \textsc{empire-3.1} nuclear
reaction code calculations to the experimental photofission cross section.

\end{abstract}


\maketitle

Photofission measurements enable selective investigation of extremely deformed
nuclear states in the light actinides and can be utilized to better understand
the landscape of the multiple-humped potential energy surface (PES) in these
nuclei. The selectivity of these measurements originates from the low and
reasonably
well-defined amount of angular momentum transferred during the photoabsorption
process. The present study is designed to investigate the PES of ${}^{238}$U
through observation of transmission resonances in the prompt photofission cross
section. A transmission resonance appears when directly-populated excited states
in the first potential minimum overlap energetically with states either in the
superdeformed (SD) 2$^\textit{nd}$ or hyperdeformed (HD) 3$^\textit{rd}$
potential minima \cite{th02,kr11}. The fission channel can thus be regarded as a
tunneling process through the multiple-humped fission barrier as the gateway
states in the first minimum decay through states in the other minima of the PES.
So far, transmission resonances have been studied primarily in
light-particle-induced nuclear reactions through charged-particle,
conversion-electron or $\gamma$-ray spectroscopy. These studies do not benefit
from the same selectivity found in photonuclear excitation and consequently they
are complicated by statistical population of the states in the 2$^\textit{nd}$
and 3$^\textit{rd}$ minima with a probability of $10^{-4} - 10^{-5}$. This
statistical population leads to a typical isomeric fission rate from the
ground-state decay of the shape isomer in the 2$^\textit{nd}$ minimum of
$\sim$1/sec. These measurements have also suffered from dominating
prompt-fission background. 

Until now, sub-barrier photofission experiments have been performed only with
brems\-strahlung photons and have determined only the integrated fission yield.
In these experiments, the fission cross section is convolved with the spectral
intensity of the $\gamma$-ray beam, resulting in a typical effective
$\gamma$-ray bandwidth $\Delta E / E$ between $4 \times 10^{-2}$ and $6 \times
10^{-2}$. These experiments observe a plateau, referred to as the ``isomeric
shelf'', in the fission cross section, resulting from competition between prompt
and delayed photofission \cite{bo78,be83}. Higher-resolution studies can be
performed at tagged-photon facilities, though only with marginal statistics, due
to the limited beam intensities realizable through tagging \cite{sa10}. This
beam intensity cannot be significantly improved beyond $\sim 10^4 ~ \gamma /
\left( \text{keV} \cdot \text{s} \right)$, since it is determined by the random
coincidence contribution in the electron-tagging process. Thus, high statistics
photofission experiments in the deep sub-barrier energy region, where cross sections
are typically as low as $\sigma$=1 nb-10 $\mu$b, cannot be performed with
tagged-photon beams. The relatively recent development of inverse-Compton
scattering $\gamma$-ray
sources, capable of producing tunable, high-flux, quasi-monoenergetic
$\gamma$-ray beams by Compton-backscattering of eV-range photons off a
relativistic electron beam, offers an opportunity to overcome previous
limitations. The present study was carried out at such a facility: the High
Intensity $\gamma$-ray Source (HI$\gamma$S) located at TUNL.
It should be emphasized that a measurement of the photofission cross section in
the deep sub-barrier energy region will be a crucial step towards a reliable
characterization of the PES, including unambiguous determination of the double-
or triple-humped nature of the surface and precise evaluation of the barrier
parameters. Next-generation Compton-backscattering $\gamma$-ray sources, such as
MEGa-ray (Lawrence Livermore National Laboratory, California, US) \cite{mega}
and ELI-NP (Bucharest, Romania) \cite{eli}, are anticipated to
provide beams with spectral fluxes of $\sim 10^6
\gamma/\left(\text{eV}\cdot\text{s}\right)$ and energy resolution of $\Delta E
\approx 1 \text{keV}$, far superior to those currently available at
HI$\gamma$S. The capabilities of these next-generation sources allow one to aim
at an identification of sub-barrier transmission resonances in the fission decay
channel with integrated cross sections down to $\Gamma \sigma \approx 0.1 ~
\text{eV} \cdot \text{b}$, whereas the present study is only sensitive to
resonances with $\Gamma \sigma \approx 10 ~ \text{eV}\cdot\text{b}$. The narrow
energy bandwidth expected for the new $\gamma$-ray beam facilities will also
allow for a significant reduction of the presently dominant background from
non-resonant processes. Thus, next-generation $\gamma$-ray sources are expected
to allow preferential population and identification of vibrational resonances in
the photofission cross section and ultimately to enable observation of the fine
structure in the isomeric shelf. This may open the perspective towards a new
era of photofission studies.

Sub-barrier photofission of $^{238}$U so far has only been studied with intense
bremsstrahlung, however, without being able to resolve any
resonances \cite{di75}. A previous $^{236}$U(p,t) measurement
showed pronounced resonance structures at excitation energies of $E^{*}=5.6-5.8$
MeV and at $E^{*}=5.15$ MeV, as well as a weaker resonance at $E^{*}=4.9$ MeV
\cite{ba74}. A whole sequence of further transmission resonances at lower
energies is expected to explain the isomeric shelf \cite{be83}, but such
resonances have not yet been observed. Furthermore, it has been found 
experimentally in several measurements on $^{234}$U \cite{kr99}, on $^{236}$U
\cite{csa05} and most recently on $^{232}$U \cite{csi09} in agreement with
older theoretical predictions \cite{cw94}, that for the uranium isotopes the HD
3$^{rd}$ potential minimum is in fact as deep as the SD 2$^{nd}$ minimum.
According to this experimental systematics, the existence of a HD 3$^{rd}$
minimum is also predicted for $^{238}$U, however, it has not yet been supported 
experimentally. On the other hand, recent calculations using a
macroscopic-microscopic model do not predict the existence of a deep 3$^{rd}$
minimum for the even-even uranium isotopes \cite{ko12,ja13}. This puzzle was
more recently addressed within a self-consistent theoretical approach, where the
conditions for the existence of HD potential minima were studied \cite{ma13}.

The aim of the present study was to measure the $^{238}$U($\gamma$,f)
cross-section at deep sub-barrier energies and to search for transmission
resonances. The experiment was performed at the HI$\gamma$S facility with its
Compton-backscattered $\gamma$-ray beam, having a bandwidth of $\Delta
E$=150-200 keV and a spectral flux of about $10^2\gamma$/(eV$\cdot$s). 
An array of parallel plate avalanche counters, consisting of 23 
electrolytically-deposited $^{238}$UO$_2$ (2 mg/cm$^2$) targets \cite{dr84}, was
used to measure the photofission cross section. Both fission fragments were
detected in coincidence to suppress the $\alpha$-particle background to an
extremely low level, which is required by the particularly low counting rates
(typically 0.1-1 Hz at E$_{\gamma}$=5 MeV). The total efficiency of the array
was estimated to be 70\% based on Ref.~\cite{dr84}.

\begin{figure}
\centering
\includegraphics[width=8.5cm]{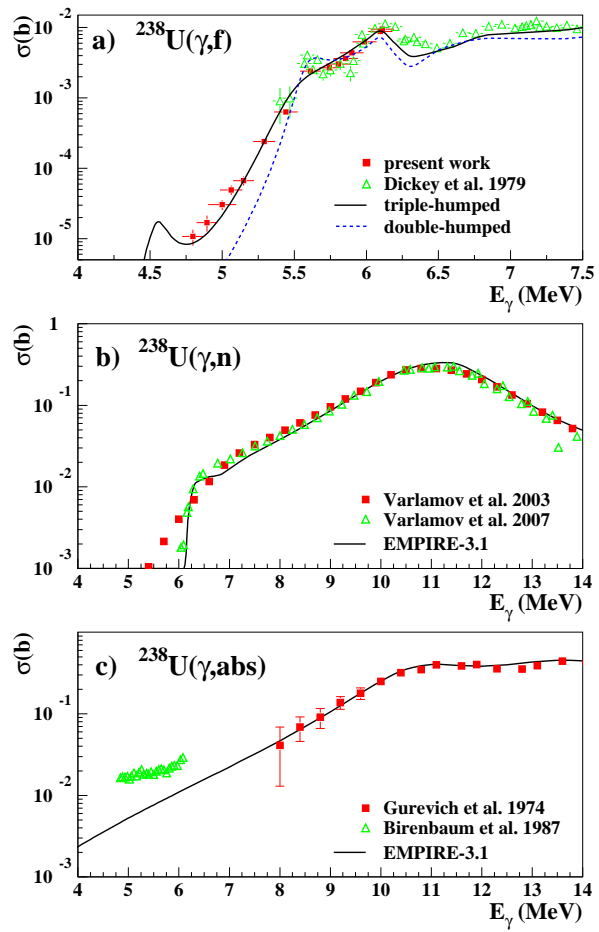}
  \caption{\label{fig:238u}(Color online) a) The measured photofission
cross-section of $^{238}$U in the $\gamma$-ray energy range of
$E_{\gamma}$=4.7-6.0 MeV. The result of the present experiment and the
experimental data of Ref.~\cite{di75} are indicated by full squares and
open triangles, respectively. b) Experimental $^{238}$U($\gamma$,n)
cross-sections of Refs.~\cite{va03} and~\cite{va07} are indicated by full
squares and open triangles, respectively. c) Total photo-absorption
cross-section of $^{238}$U as a function of the $\gamma$-ray energy. The
experimental data from Ref.~\cite{gu76} and Ref.~\cite{bi87} are
indicated by full squares and open triangles, respectively. In all panels, the
cross sections calculated using \textsc{empire}-3.1, as discussed in the text,
are shown as black lines; the calculations in panel b) and c) assume a triple-humped barrier structure, however, without influencing the resulting cross sections.}
\end{figure}

The present experimental photofission cross-section of $^{238}$U as a function
of the
$\gamma$-ray energy is shown in Fig.~\ref{fig:238u}a, along with the
experimental data of Ref.~\cite{di75}. Near the top of the barrier the two data
sets are in a good agreement. The present data are extended by about an order of
magnitude in cross section to the deep sub-barrier region down to
$E_{\gamma}$=4.7 MeV. A clear transmission resonance has been observed at
$E_{\gamma}$=5.6 MeV, which is consistent with the observation
of Ref.~\cite{di75}. A slight deviation from the exponential slope of the cross
section indicates the existence of a resonance at $E_{\gamma}$=5.1 MeV, however,
with only a limited resonance signal contrast due to the moderate bandwidth of
the $\gamma$-ray beam. 

For the theoretical evaluation of the present $^{238}$U photofission
experimental data, we performed nuclear reaction code calculations using the
\textsc{empire}-3.1 code \cite{he07}. Within the code, the fission transmission
coefficients are calculated using the Hill-Wheeler formalism \cite{hi53},
followed by Hauser-Feshbach statistical model calculations \cite{ha52}, allowing
the fission channel to compete with emission of particles and photons. The triple-humped fission barrier
parameters of $^{238}$U were extracted by tuning the inputs to these
calculations and comparing the resulting predictions of the photofission cross
section to the experimental data.

The reliability of the code was tested and the relevant model parameters were
adjusted using calculations of the total photo-absorption cross section
$\sigma_{\gamma,\text{abs}}$ and experimental ($\gamma$,n) cross section data.
First, $\sigma_{\gamma,\text{abs}}$ had to be determined and checked against
existing experimental data. In the present evaluation, the modified
Lorentzian parameterization (MLO) was chosen for the $\gamma$-ray strength 
function. Although the experimental data of Ref.~\cite{gu76} are quite well
reproduced (solid line in Fig.~\ref{fig:238u}c), the experimental results of
Ref.~\cite{bi87} are underestimated at lower energies. Yet, we have not
attempted to tune the MLO parameters to reproduce the experimental data. The
parameterization used is based on a global fit of experimental data 
over a wide range of isotopes and excitation energies. Attempts to reproduce
this dataset would have a drastic impact on the competing reaction channels,
leading (especially for fission) to unphysical parameters. The photo-absorption
cross sections of Ref.~\cite{bi87} were inferred from the measured
energy-averaged, angle-integrated photon elastic-scattering cross 
sections $\sigma_{\gamma\gamma}$, employing a complex analysis technique
described in details in Ref.~\cite{ax70}. In such an analysis, the measured
values are renormalized by an energy-dependent factor to obtain the corrected
photo-absorption cross section $\sigma_{\gamma,\text{abs}}$. Our calculated
values are located between the measured $\sigma_{\gamma\gamma}$ and the
corrected $\sigma_{\gamma,\text{abs}}$ cross sections, perhaps 
indicating systematic uncertainties in the aforementioned analysis.

The transmission coefficients for the particle emission were determined using
the global optical parameter set of Ref.~\cite{ca08}. The level density
parameters were taken from the enhanced generalized super-fluid model
\cite{da94}, adjusted to fit the discrete
level scheme of $^{238}$U. Those were taken from the most recent reference input
parameter library (RIPL3). In the code, the optical model for fission
\cite{bh80,si06,si08} is applied to calculate the fission transmission
coefficients. For comparison, both triple- and double-humped fission barriers
were used in the calculations.

\begin{table}
  \caption{\label{tab:param2}Double-humped fission barrier parameters of $^{238}$U (in MeV) 
      used in the calculations. The resulting photofission cross section is
indicated in Figure \ref{fig:238u}a by the 
      dashed line.} 
\begin{ruledtabular} 
  \begin{tabular}{cccccccccc} 
     $E_A$ & $E_{II}$ & $E_B$ & $\hbar\omega_A$ & $\hbar\omega_{II}$ & $\hbar\omega_B$ \\ 
\hline
    6.3$\pm$0.2 & 2.0$\pm$0.2 & 5.65$\pm$0.20 & 1.1$\pm$0.1 & 1.0$\pm$0.1 &
0.6$\pm$0.1
  \end{tabular}\end{ruledtabular} 
\end{table}

\begin{table*}
  \caption{\label{tab:param1}Triple-humped fission barrier parameters of
$^{238}$U 
          (all in MeV) used in the calculation, represented by the solid line
in 
          Figure \ref{fig:238u}a.} 
\begin{ruledtabular} 
  \begin{tabular}{cccccccccc} 
    $E_A$ & $E_{II}$ & $E_B$ & $E_{III}$ & $E_C$ & $\hbar\omega_A$ &
$\hbar\omega_{II}$ & 
                   $\hbar\omega_B$  &$\hbar\omega_{III}$ & $\hbar\omega_C$\\ 
\hline
     4.3$\pm$0.2 & 2.05$\pm$0.20 & 5.6$\pm$0.2 & 3.6$\pm$0.2 & 5.6$\pm$0.2 
                 & 0.4$\pm$0.1 & 1.0$\pm$0.1 & 0.7$\pm$0.1 & 1.0$\pm$0.1 &
0.7$\pm$0.1
  \end{tabular}
\end{ruledtabular} 
\end{table*}

The parameters of the double-humped fission barrier were taken from the RIPL3
library and were slightly adjusted to achieve a better description of the
present data. In Figure \ref{fig:238u}a, the dashed line shows the calculated
($\gamma$,f) cross-section using the parameters listed in Table
\ref{tab:param2}. The triple-humped barrier parameters were adjusted to best describe
the experimental photofission and ($\gamma$,n) cross sections
over the entire energy range. In Figure \ref{fig:238u}a, the solid line
represents the best description with the parameters of Table \ref{tab:param1}
used in the calculation. The calculated ($\gamma$,n) cross-sections are shown as
the solid line in Fig.~\ref{fig:238u}b together with the available experimental
data \cite{va03,va07}. The calculated and the experimental values are in a fair
agreement. The uncertainties of the barrier parameters were estimated to be 200
keV for the barrier heights and 100 keV for the curvature parameters. 

The present model is capable of reproducing the sub-barrier fission resonances
empirically, while at higher excitation energies it naturally provides the same
results for the fission barrier penetration as the classical models. Since photofission occurs only through
the giant dipole resonance, as a good approximation, only negative parity
states are important for $\gamma$-ray-induced fission and contributions from M1 
and E2 excitations are very small. Positive parity states were involved in the calculations only to
achieve consistency with the neutron-induced fission cross sections (e.g.
n+$^{238}$U, where $^{238}$U is involved in the second chance fission).

\begin{figure}[!]
\centering
\includegraphics[width=8.5cm]{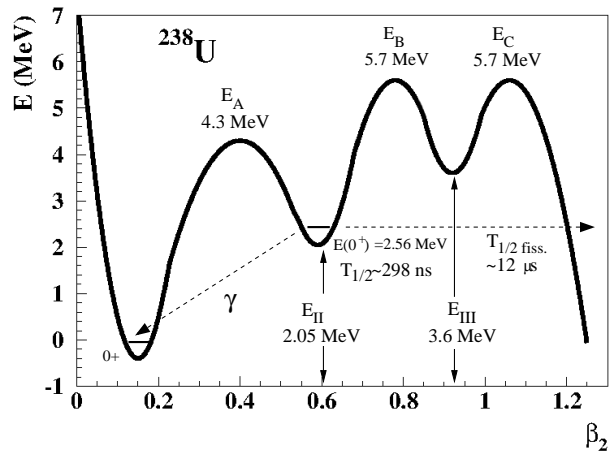}
\caption{\label{fig:barrier} The triple-humped fission barrier of $^{238}$U
as determined in the present study, using the parameters listed in Table
\ref{tab:param1}. The half-life of the isomeric ground state at $E$=2.56 MeV and
the partial isomeric fission half-life are also indicated.}
\end{figure}

The experimental data of the present experiment could be reproduced
dramatically better with a calculation assuming a triple-humped fission barrier
than with a double-humped one. When using a triple-humped barrier, an additional
resonance at $E_{\gamma}$=4.6 MeV had to be included in the calculations.
Experimental evidence for the existence of such a resonance would fully confirm
our present theoretical interpretation. It is also evident that the existing
($\gamma$,f) and ($\gamma$,n) experimental data suffer from large uncertainties.
It would be highly important to improve the $\gamma$-ray beam energy
bandwidth and the energy density of data points (requiring a higher $\gamma$-ray
beam intensity), in order to explore a full set of deep sub-barrier fission
resonances.

\begin{figure}[!]
\centering
\includegraphics[width=8.5cm]{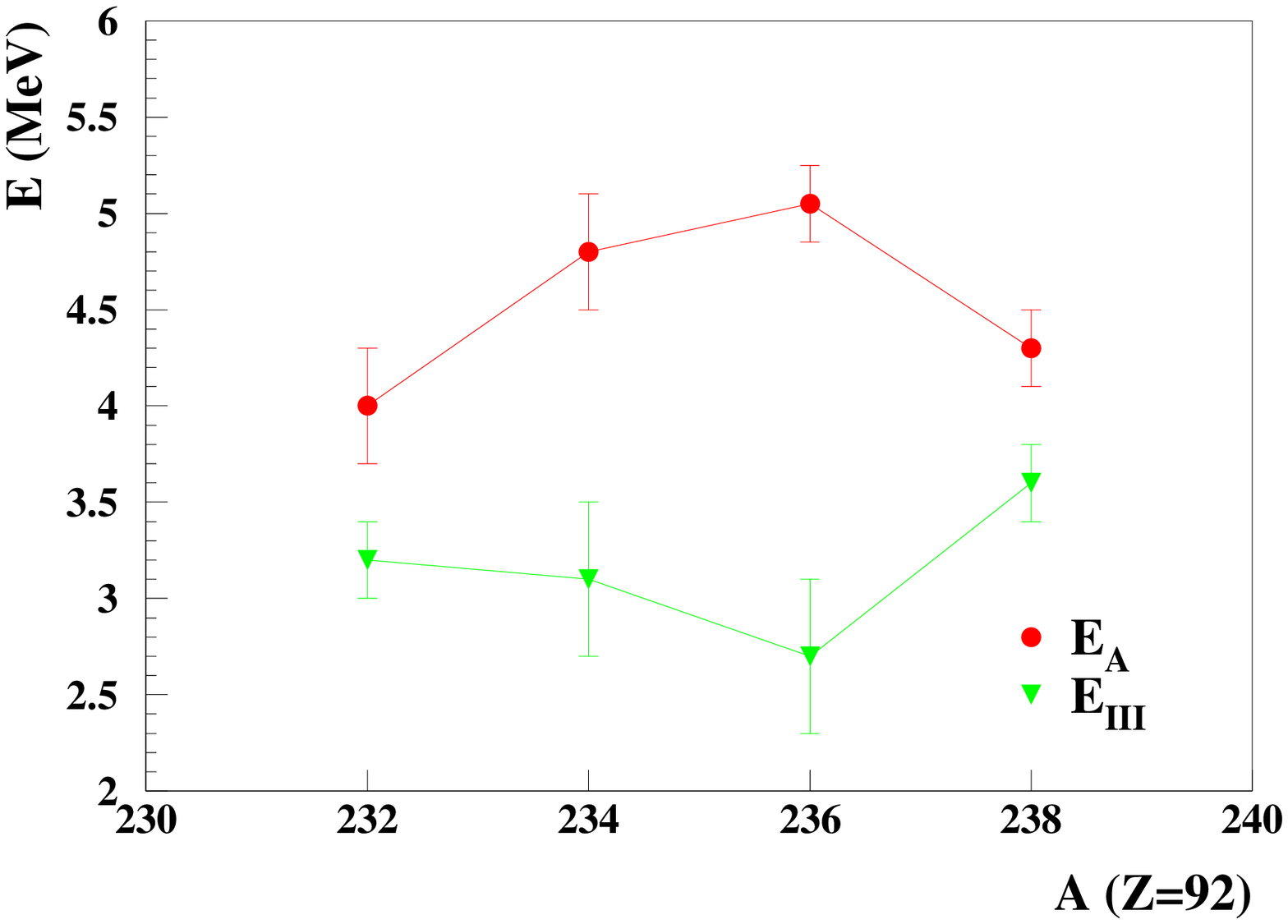}
\caption{\label{fig:syst} (Color online) The height of the inner barrier
$E_{\text{A}}$ and the depth of the third minimum $E_{\text{III}}$ for
even-even uranium isotopes, shown as red circles and green
triangles, respectively. The experimental data for $^{232}$U, $^{234}$U
and $^{236}$U were taken from Refs.\cite{kr99,csa05,csi09}.}
\end{figure}

The present results on the fission barrier parameters of $^{238}$U supplement
the previous findings on the systematics of the barrier parameters of the
uranium isotopes \cite{csi09}. Fig.~\ref{fig:syst} shows the present results on
$^{238}$U together with previous experimental results on $^{232}$U
\cite{csi09}, $^{234}$U \cite{kr99} and $^{236}$U \cite{csa05}. A
reversal of the trends followed by the lighter uranium isotopes for the height
of the inner barrier E$_A$ and the depth of the third minimum
(expressed by E$_{III}$), respectively, as a function of the neutron number, is
observed. For $^{238}$U, the data suggests a decreasing barrier height $E_A$ and
a decreased depth of the third minimum. Moreover, the particularly low values of
the curvature parameters derived from the present data, especially the one for
the inner barrier ($\hbar\omega_A$=0.4
MeV), may suggest a need for reconsideration of the well-accepted approximation
of the fission barrier with a harmonic oscillator potential curve. An
anharmonic, ``tower-like'' potential, originally suggested by Bowman
\textit{et al}. decades ago \cite{bo75}, would better approximate the potential
landscape
determined from the current data.

In summary, we measured the photofission cross-section of $^{238}$U in the
$\gamma$-ray energy region of $E$=4.7-6.0 MeV with the monochromatic,
high-brilliance, Compton-backscattered $\gamma$-ray beam of the HI$\gamma$S
facility. With the significantly higher intensity of the beam, when comparing to
a tagged-photon facility, the cross-section could be measured at deep 
sub-barrier energies. \textsc{empire}-3.1 reaction code calculations were
performed to extract the fission barrier parameters of $^{238}$U. Our present
results on the fission barrier of $^{238}$U support a deep 3$^{rd}$ minimum with
$E_{\text{III}}$=3.6 MeV, a low inner barrier height $E_{\text{A}}$=4.3 MeV and outer
barrier heights of $E_{\text{B}}$=5.7 MeV and $E_{\text{C}}$=5.7 MeV. Though in
line with the extensive body of experimental evidence for deep third
potential minima in uranium isotopes acquired over the last 15 years, this
result is in disagreement with recent calculations of Ref.~\cite{ko12}, a puzzle
that still needs to be resolved. Indications of predicted resonance structures
have also been observed, however,
with moderate amplitudes. The results indicate the need for further
investigations at lower $\gamma$-ray energies and using smaller-bandwidth,
higher-intensity $\gamma$-ray beams. ELI-NP, MEGa-ray, and other next-generation
$\gamma$-ray sources will enable such measurements.

The work has been supported by the DFG Cluster of Excellence ``Origin and
Structure of the Universe'', the Hungarian OTKA Foundation No. K106035 and the
USDOE grant No. DE-FG02-97ER410U.

\end{document}